\documentclass{moriond}

\usepackage{amsmath,amssymb,cite}

\newcommand{\Order}{\mathcal{O}}

\newcommand{\GeV}{\,\text{GeV}}

\newcommand{\Lagr}{\mathcal{L}}
\newcommand{\beq}{\begin{equation}}
\newcommand{\eeq}{\end{equation}}
\renewcommand{\Re}{\text{Re}\,}
\renewcommand{\Im}{\text{Im}\,}
\newcommand{\Br}{\text{Br}}
\newcommand{\ecm}{e\,\text{cm}}

\newcommand{\Photo}{}

\begin{document}
\vspace*{4cm}
\title{COMBINED EXPLANATIONS OF $\boldsymbol{(g-2)_{\mu}}$, $\boldsymbol{(g-2)_e}$ AND\\[0.2cm] IMPLICATIONS FOR A LARGE MUON EDM}

\author{A.\ CRIVELLIN$^{1,2}$ and M.\ HOFERICHTER$^3$}

\address{$^1$Paul Scherrer Institut, CH--5232 Villigen PSI, Switzerland\\
$^2$Physik-Institut, Universit\"at Z\"urich, Winterthurerstrasse 190, CH--8057 Z\"urich, Switzerland\\
$^3$Institute for Nuclear Theory, University of Washington, Seattle, WA 98195-1550, USA}

\maketitle

\abstracts{We consider possible beyond-the-Standard-Model (BSM) effects that can accommodate both the long-standing tension in the anomalous magnetic moment of the muon, $a_\mu=(g-2)_\mu/2$,
as well as the emerging $2.5\sigma$ deviation in its electron counterpart, $a_e=(g-2)_e/2$. After performing an EFT analysis, we consider BSM physics realized above the electroweak scale and 
find that a simultaneous explanation becomes possible in models with chiral enhancement. However, this requires a decoupling of the muon and electron BSM sectors to avoid the 
strong constraints from $\mu\to e\gamma$. In particular, this decoupling implies that there is no reason to expect the muon electric dipole moment (EDM) $d_\mu$ to be correlated 
with the electron EDM $d_e$, avoiding the very stringent limits for the latter. While some of the parameter space for $d_\mu$ favored by $a_\mu$ could be tested at the $(g-2)_\mu$ experiments 
at Fermilab and J-PARC, a dedicated muon EDM experiment at PSI would be able to probe most of this region.}

\section{Status of lepton dipole moments}

The experimental value of the muon $g-2$~\cite{Bennett:2006fi}
\beq
a_\mu^\text{exp}=116,\!592,\!089(63)\times 10^{-11}
\eeq
differs from the SM prediction at the level of $3$--$4\sigma$, for definiteness we take~\cite{Crivellin:2018qmi}
\beq
\label{Delta_amu}
\Delta a_\mu=a_\mu^\text{exp} - a_\mu^\text{SM}\sim 270(85)\times 10^{-11}
\eeq
as an estimate of the current status. Recent advances in corroborating and improving the SM prediction include hadronic vacuum polarization~\cite{Jegerlehner:2017lbd,Davier:2017zfy,Borsanyi:2017zdw,Blum:2018mom,Keshavarzi:2018mgv,Colangelo:2018mtw,Davies:2019efs,Gerardin:2019rua}, 
hadronic light-by-light scattering~\cite{Colangelo:2015ama,Green:2015sra,Blum:2016lnc,Colangelo:2017qdm,Colangelo:2017fiz,Blum:2017cer,Hoferichter:2018dmo,Hoferichter:2018kwz,Gerardin:2019vio}, and 
higher-order hadronic corrections~\cite{Kurz:2014wya,Colangelo:2014qya}. The release of first results from the Fermilab experiment~\cite{Grange:2015fou} is highly anticipated, while a complementary strategy based on ultracold muons is being pursued at J-PARC~\cite{Saito:2012zz}, see also Ref.~\cite{Gorringe:2015cma}.

For the electron, the direct measurement~\cite{Hanneke:2008tm}
\beq
a_e^\text{exp}=1,\!159,\!652,\!180.73(28)\times 10^{-12}
\eeq
displays a tension with the SM prediction 
\beq
a_e^\text{SM}=1,\!159,\!652,\!181.61(23)\times 10^{-12}
\eeq
at the level of $2.5\sigma$
\beq
\label{Delta_ae}
\Delta a_e=a_e^\text{exp} - a_e^\text{SM}=-0.88(36)\times 10^{-12}.
\eeq
The key input for the SM prediction is the fine-structure constant $\alpha$, now known from atomic interferometry in Cs~\cite{Parker:2018vye} at a level that matches (and even slightly exceeds) the precision of the direct electron
$g-2$ measurement. This direct translation of improved input for $\alpha$ to $a_e^\text{SM}$ is possible thanks to the semi-analytical calculation~\cite{Laporta:2017okg}, which removes the uncertainty in the four-loop QED coefficient altogether, as well as to the improved numerical calculation of the five-loop coefficient~\cite{Aoyama:2017uqe}. In fact, theory uncertainties are now at a level sufficient for an improvement by another order of magnitude on the experimental side.  

For the EDMs the current best limits are~\cite{Bennett:2008dy,Baron:2013eja,Andreev:2018ayy}
\beq
\label{dmulimit}
|d_\mu|<1.5\times 10^{-19}\ecm,\qquad |d_e|<1.1\times 10^{-29}\ecm\qquad 90\% \,\text{C.L.}
\eeq
Assuming minimal flavor violation (MFV), i.e.\ a linear scaling in the mass, one would thus expect 
\beq
\label{dmu_MFV}
|d_\mu^\text{MFV}|=\frac{m_\mu}{m_e}|d_e|<2.3\times 10^{-27}\ecm,
\eeq
which is beyond the reach of any foreseeable experiment. However, 
the MFV paradigm is being challenged by the anomalies observed in semileptonic $B$ meson decays (see Ref.~\cite{Crivellin:2018gzw} for a recent review)
and also cannot account for both $g-2$ tensions simultaneously. There is therefore strong motivation to study scenarios that go beyond the MFV picture.

\begin{figure}[t!]
	\centering
	\includegraphics[height=4.6cm]{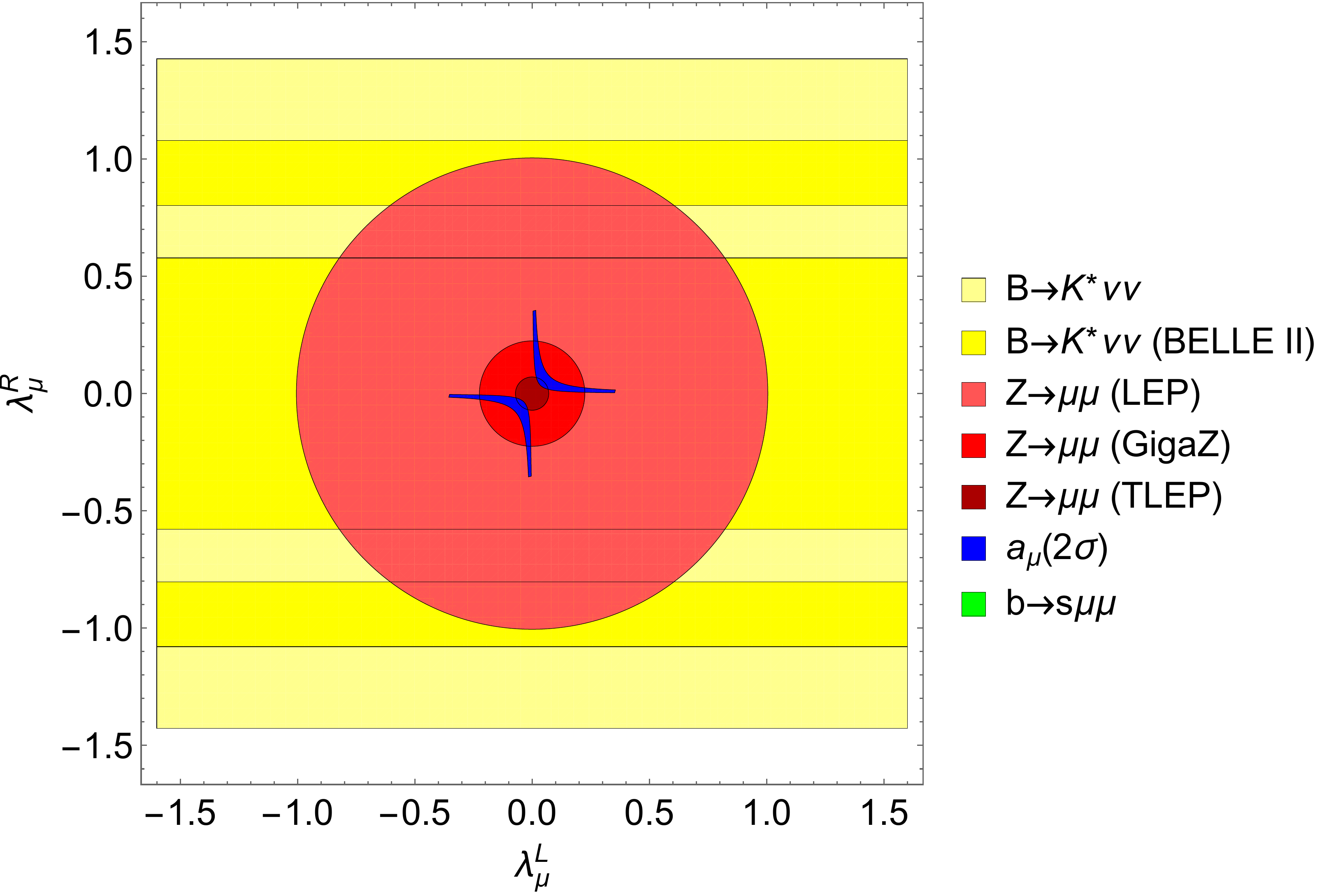}
	\includegraphics[height=4.2cm]{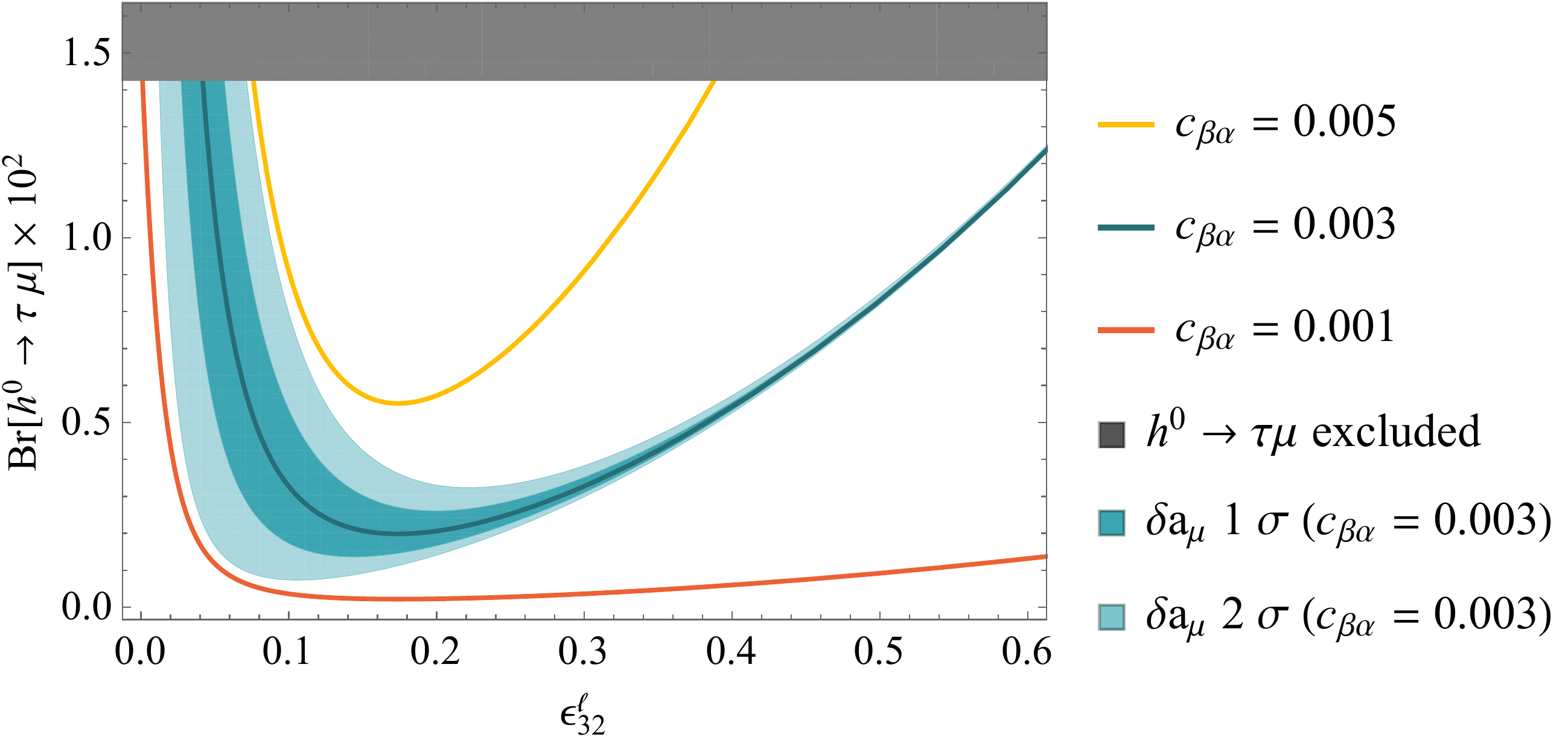}
	\caption{Left: Allowed regions in the $\lambda^L_\mu$--$\lambda^R_\mu$ plane ($\lambda^{L(R)}_\mu$ is the coupling of the LQ to left-handed (right-handed) muons and the top quark) from current and future experiments for the $SU(2)$ singlet LQs $\Phi_1$ with $M=1\,$TeV. Right: Prediction for the decay of the SM-like Higgs boson $h\to \tau\mu$ as a function of $\varepsilon^{\ell}_{32}$ (the coupling of the second Higgs doublet to $\tau_L\mu_R$) under the assumption that $\varepsilon^{\ell}_{23}$ (the $\tau_R\mu_L$ coupling) is chosen in such a way that $a_\mu$ is explained. We used $M_{H^+}=400\GeV$, $M_{H_0}=250\GeV$, and $M_{A_0}=300\GeV$. For $c_{\beta\alpha}=0.003$ (the mixing among the $CP$ even neutral Higgses) the whole $2\sigma$ region to explain $a_\mu$ is shown, while for $c_{\beta\alpha}=0.001$ and $c_{\beta\alpha}=0.005$ only the predictions for the central value of $a_\mu$ are depicted. Note that $h\to\tau\mu$ enforces a tight alignment of the Higgs sector.
	Figure taken from Refs.~\cite{ColuccioLeskow:2016dox,Crivellin:2019dun}.}
	\label{fig:LQ}
\end{figure}

In an effective field theory (EFT) approach, $g-2$ and EDM correspond to the real and imaginary part of the Wilson coefficient $c^{\ell\ell}_{R}$ of the operator $\bar{\ell}_{f}
\sigma_{\mu \nu} P_{R} \ell_{i} F^{\mu \nu}$, respectively. The off-diagonal terms describe lepton-flavor-violating processes, most notably in this context $\mu\to e\gamma$.
The current experimental limits on the EDMs together with the $g-2$ deviations in Eqs.~\eqref{Delta_amu} and~\eqref{Delta_ae} thus limit the phases according to
\beq
\label{limits_phase}
\bigg|\frac{\Im c^{ee}_{R}}{\Re c^{ee}_{R}}\bigg|\lesssim 6\times 10^{-7},\qquad
\bigg|\frac{\Im c^{\mu\mu}_{R}}{\Re c^{\mu\mu}_{R}}\bigg|\lesssim 600,
\eeq
extremely small for the electron, but largely unconstrained for the muon. In addition, the EFT analysis 
implies that BSM scenarios fulfilling
$c^{e\mu}_{R}=\sqrt{c^{ee}_{R}c^{\mu\mu}_{R}}$ are excluded since the resulting
\beq
\Br[\mu \to e \gamma]  = \frac{\alpha m_\mu ^2}{16 m_e\Gamma _\mu }|\Delta a_\mu \Delta a_e|\sim 8\times 10^{-5}
\label{muegamma}
\eeq
violates the MEG bound~\cite{TheMEG:2016wtm}
\beq
\Br[\mu \to e \gamma]<4.2\times 10^{-13}\qquad 90\% \,\text{C.L.}
\eeq
by $8$ orders of magnitude. This demonstrates the necessity of carefully decoupling the electron and muon BSM sectors.

\section{Possible BSM explanations}

The deviation from the SM expectation in Eq.~\eqref{Delta_amu} is large, with a central value nearly twice the size of the electroweak SM contribution~\cite{Gnendiger:2013pva}. For that reason, any viable BSM mechanism needs to involve some sort of enhancement. Such an enhancement can be produced by light particles, but dark (axial) photons are problematic because they necessarily lead to a positive (negative) sign and therefore increase the tension in either $(g-2)_e$ or $(g-2)_\mu$. More complicated constructions based on a light scalar and the interplay of one- and two-loop processes are possible~\cite{Davoudiasl:2018fbb}, but lead to a real Wilson coefficient and thus a vanishing EDM.

Here, we concentrate on scenarios with BSM physics realized above the electroweak scale. In this case, a possible enhancement mechanism is related to the chirality flip, which can be provided by a new heavy fermion instead of $m_\ell$. Examples for such a chiral enhancement are $\tan\beta$ in the MSSM (see e.g.~\cite{Stockinger:2006zn} for a review), $m_t/m_\ell$ in leptoquark (LQ) models~\cite{Djouadi:1989md,ColuccioLeskow:2016dox}, or $m_\tau/m_\mu$ in two-Higgs-doublet (2HDM) models~\cite{Crivellin:2019dun,Abe:2019bkf}. In general, in any model with chiral enhancement, $c^{\ell\ell}_{R}$ can be complex with an a priori unconstrained phase, so that the resulting muon EDM can become sizable. As an example for models with such chirally-enhanced effects, the situation for the scalar singlet LQ $S_1$~\cite{ColuccioLeskow:2016dox} and 2HDMs with $\tau$--$\mu$ flavor violations~\cite{Crivellin:2019dun} is shown in Fig.~\ref{fig:LQ}. Here, one can see that it is possible to test (indirectly) the LQ explanation with future $Z\to\mu^+\mu^-$ measurements at the FCC-ee, while the 2HDM explanation with $\tau$--$\mu$ flavor violations leads to $h\to\tau\mu$. Note that both of these models can explain $a_\mu$ and have a free phase, so that $d_\mu$ can be sizable. However, since these are single-mediator models, they are subject to the constraint in Eq.~\eqref{muegamma} and cannot provide a simultaneous explanation of $a_e$.

\begin{figure}[t!]
	\centering
		\includegraphics[height=6cm]{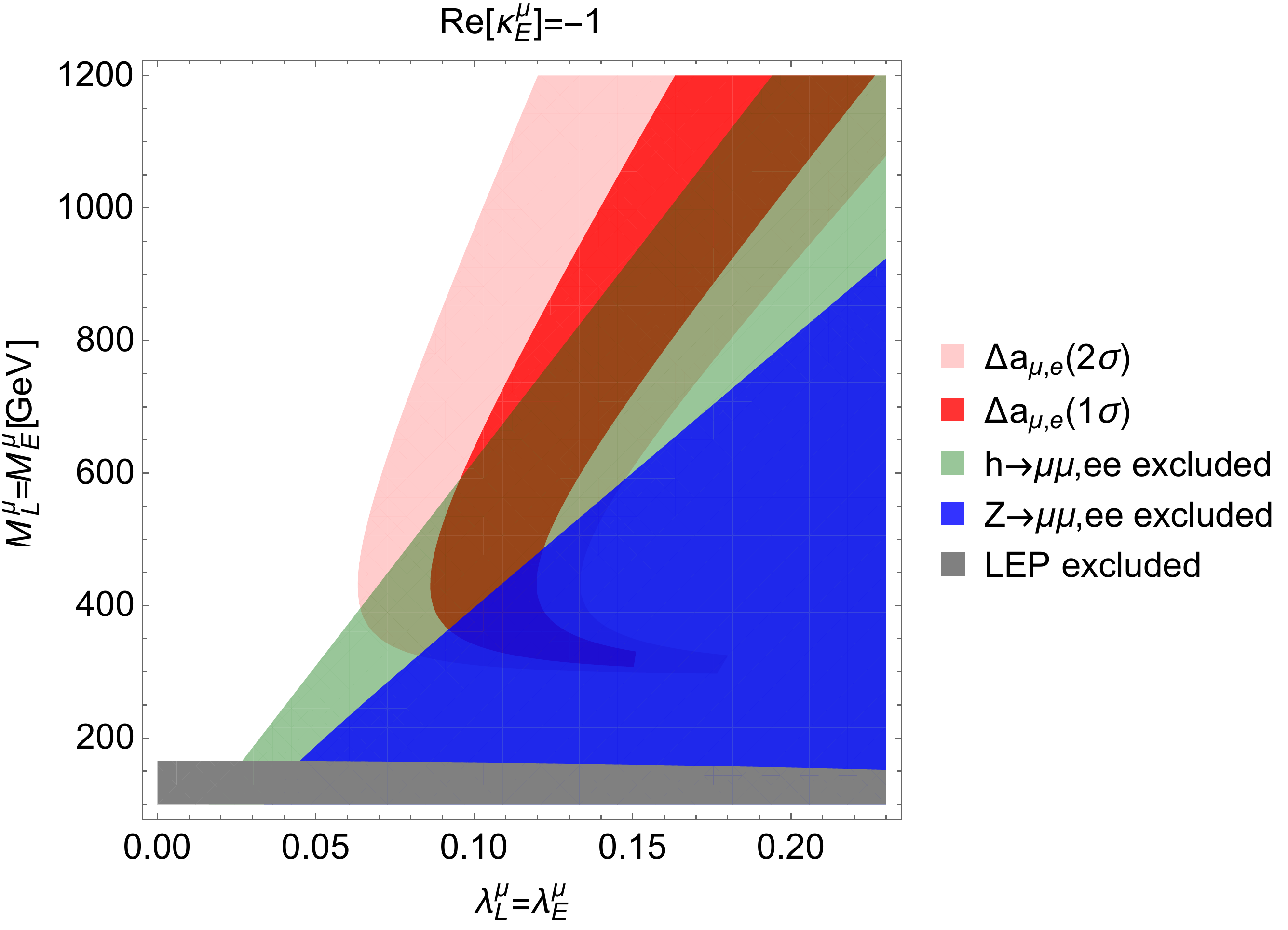}
		\includegraphics[height=6cm]{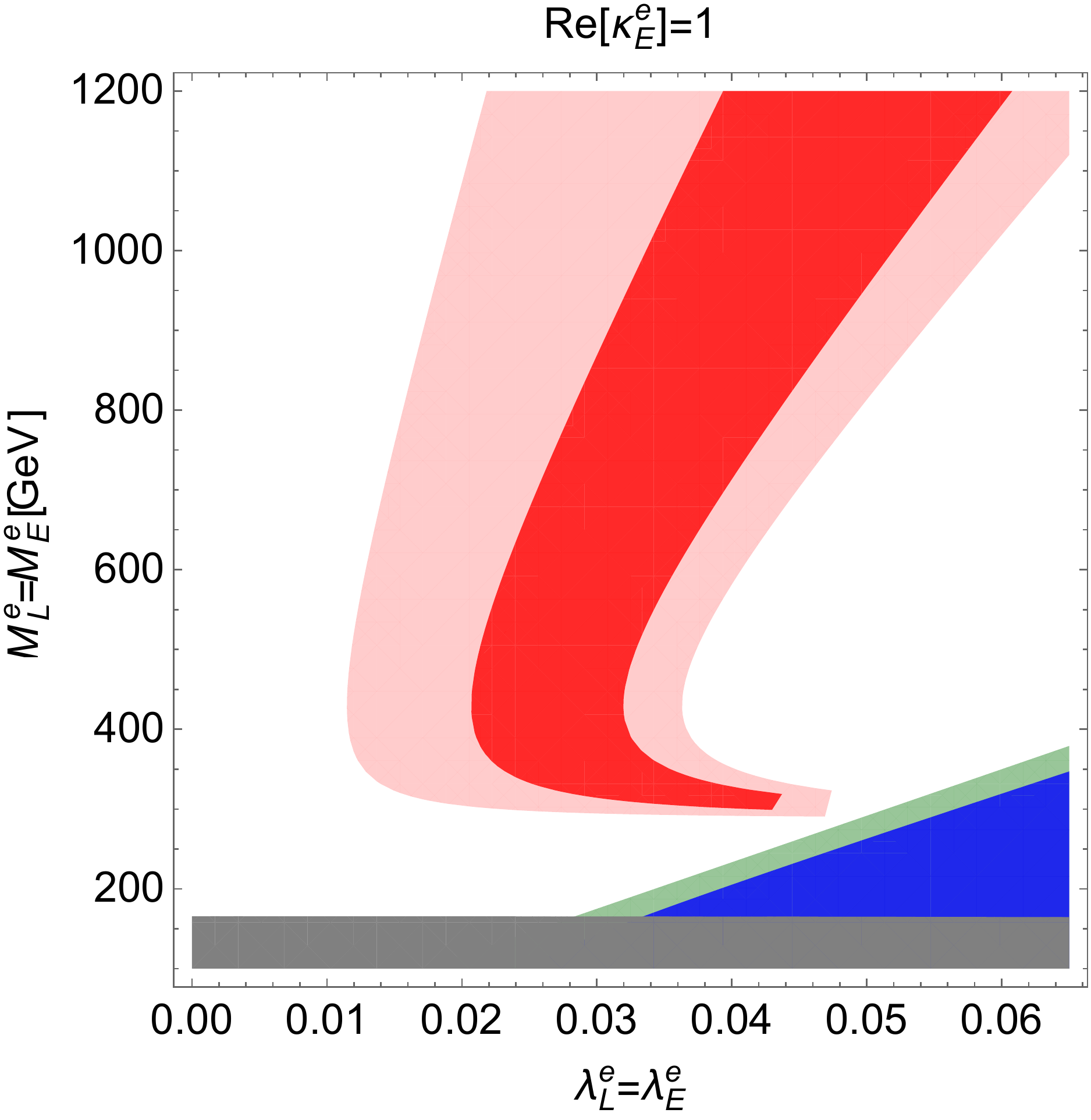}
	\caption{Allowed regions of $a_\ell$ in the $\lambda_E=\lambda_L$--$M_E=M_L$ plane for $\kappa_L=0$ and $\kappa_E=\mp1$ for muon (left) and electron (right), with bounds from
	$\sigma(h\to\mu^+\mu^-)/\sigma(h\to\mu^+\mu^-)_\text{SM}=0\pm 1.3$~\cite{Patrignani:2016xqp,Khachatryan:2016vau,Aaboud:2017ojs},
	$\sigma(h\to e^+e^-)/\sigma(h\to e^+e^-)_\text{SM}<3.7\times 10^5$~\cite{Khachatryan:2014aep}, $Z\to\ell\ell$~\cite{Patrignani:2016xqp,ALEPH:2005ab},
	and direct searches for new heavy charged leptons~\cite{Achard:2001qw}. Figure taken from Ref.~\cite{Crivellin:2018qmi}.}
	\label{fig:aellHiggs}
\end{figure}

In Ref.~\cite{Crivellin:2018qmi} we constructed a UV complete model with vector-like leptons
\begin{align}
	\Lagr_M&=  - {M_L}{{\bar L}_L}{L_R} - {M_E}{{\bar E}_L}{E_R} - {\kappa _L}{{\bar L}_L}H{E_R} - \kappa _E^{}{{\bar L}_R}H{E_L} + \text{h.c.}
\end{align}
and either the SM Higgs ($H$) or a new heavy scalar ($\phi$) coupling the vector-like leptons to the SM ones ($\ell$)
\begin{align}
 \Lagr_{H}&=   - {\lambda _L}{{\bar L}_L}{\ell_R}H - {\lambda _E}{{\bar E}_R}\tilde H{\ell _L} + \text{h.c.},
 \end{align}
 and similarly for $\phi$.
 This setup leads to a chiral enhancement by $\kappa_{L,R}v/m_\ell$, with $v$ denoting the SM Higgs vev. As illustrated in Fig.~\ref{fig:aellHiggs}, the variant with the SM Higgs as the only scalar works well for the electron,
 while for the muon the parameter space is already significantly constrained. A minimal model can be obtained when the electron $g-2$ is explained via a loop involving the SM Higgs, while for the muon
 a new heavy scalar contributes. Furthermore, $\mu\to e\gamma$ transitions can be avoided by an Abelian flavor symmetry such as $L_\mu-L_\tau$. In all these cases there are no correlations between $c_R^{ee}$ and $c_R^{\mu\mu}$,
 so that the phase of $c_R^{\mu\mu}$ is not constrained by $|d_e|$ and thus $|d_\mu|$ can be sizable.

\section{Implications for the muon EDM}

\begin{figure}[t!]
\centering
\includegraphics[height=6cm]{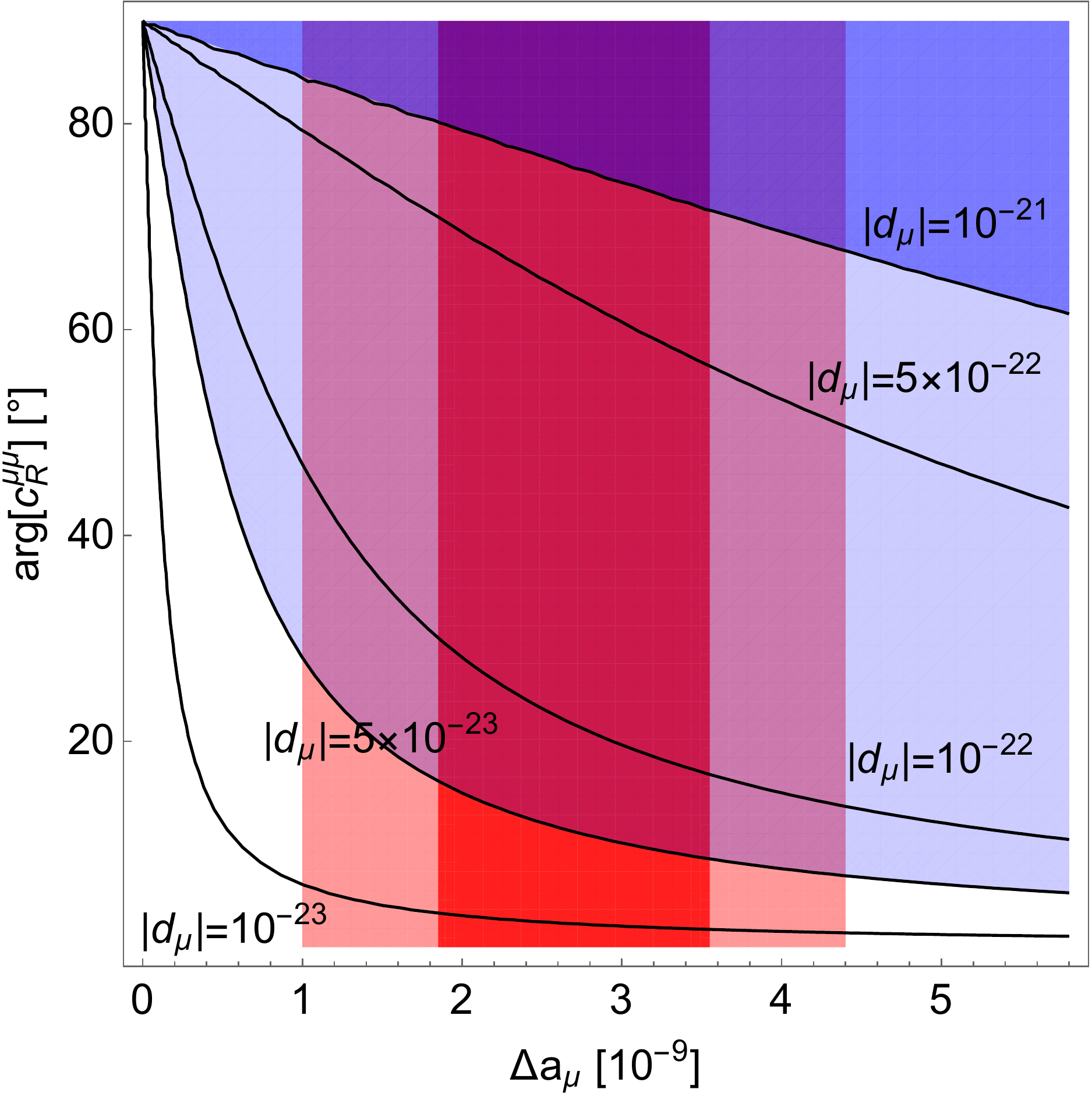}\qquad
\includegraphics[height=6cm]{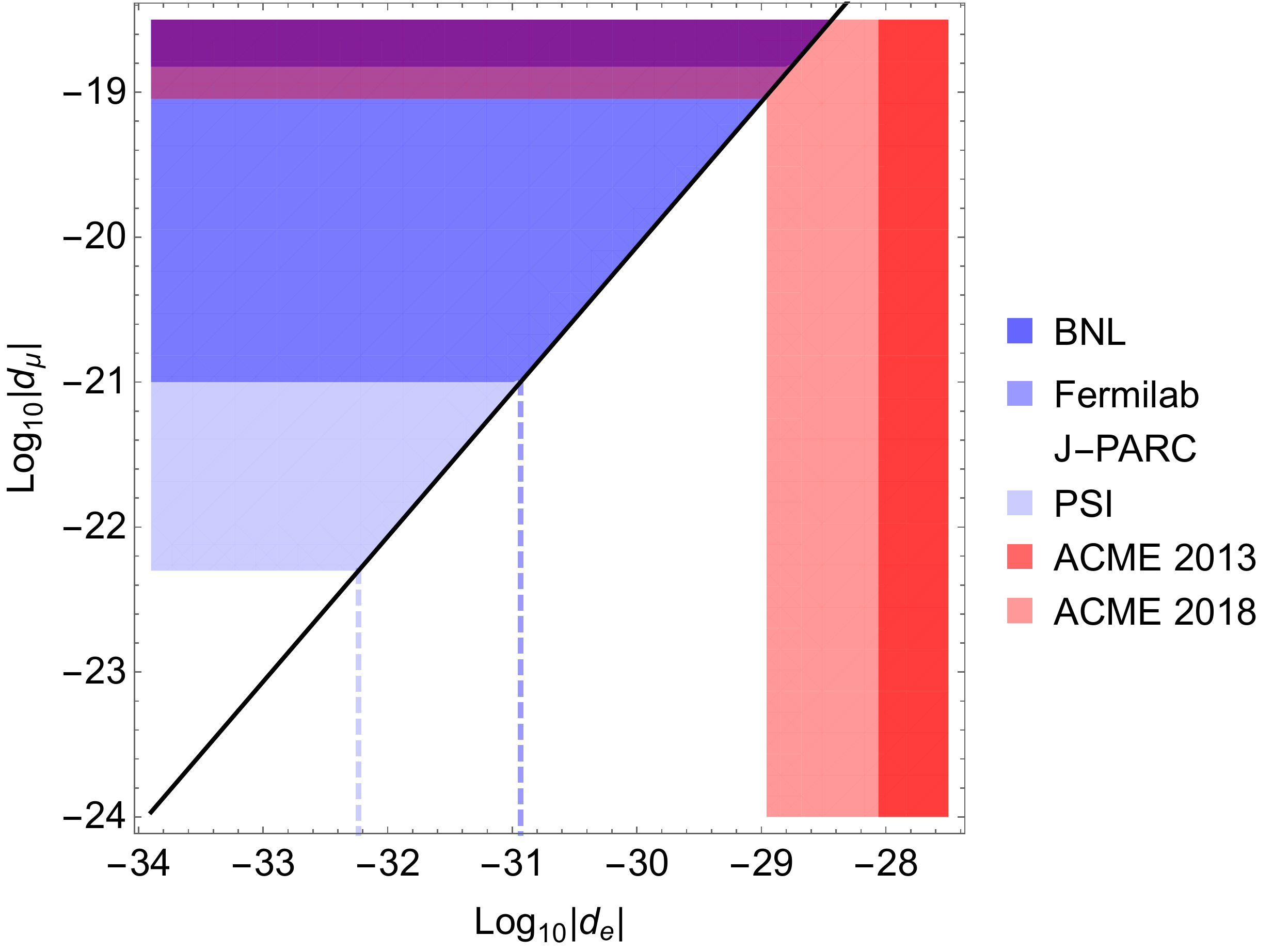}
\caption{Left: Contour lines defining the muon EDM (in units of $\ecm$) as a function of $\Delta a_\mu$ and the phase of the Wilson coefficient $c_R^{\mu\mu}$. The red regions are currently preferred by the measurement of $a_\mu$ and the blue regions are the expected sensitivity of the Fermilab/J-PARC (dark blue) and the proposed PSI experiment (light blue). The limit on the phase derived from the current limit for $|d_\mu|$ is so close to $90^{\circ}$ that it is not visible in the plot.
Right: Present and future direct limits on $|d_\mu|$ from BNL~\cite{Bennett:2008dy} (dark blue), see Eq.~\eqref{dmulimit}, Fermilab/J-PARC (blue), and the proposed PSI experiment (light blue). The dark red and light red regions refer to the ACME 2013~\cite{Baron:2013eja} and ACME 2018~\cite{Andreev:2018ayy} limits on $|d_e|$, respectively, where the latter provides an indirect bound on $|d_\mu|$ slightly 
better than the BNL direct bound. The blue dashed lines indicate the limits on $|d_e|$ that would be required to match the anticipated direct limits from Fermilab/J-PARC and PSI.
The black line defines the relation~\eqref{indirect_bound}, with the upper-left half referring to limits on $|d_\mu|$ and the lower-right to limits on $|d_e|$.
Figure taken from Ref.~\cite{Crivellin:2018qmi}.}
\label{fig:sensitivity}
\end{figure}

Even when the muon and electron BSM sectors are decoupled in a particular model, there is still a contribution of the muon EDM operator to the electron EDM via a three-loop diagram~\cite{Grozin:2009jq}, leading to the indirect constraint
\begin{align}
\label{indirect_bound}
|d_\mu|\leq \Bigg[\bigg(\frac{15}{4}\zeta(3)-\frac{31}{12}\bigg)\frac{m_e}{m_\mu}\bigg(\frac{\alpha}{\pi}\bigg)^3\Bigg]^{-1}|d_e|
\leq 0.9\times 10^{-19}\ecm\qquad 90\% \,\text{C.L.}
\end{align}
Currently, the limit on $|d_e|$ is so much more stringent than the direct one on $|d_\mu|$ that it suffices to overcome the three-loop suppression. In the future, however, the $(g-2)_\mu$ experiments at Fermilab and J-PARC
are expected to probe the muon EDM at a level of $10^{-21}\ecm$, which could be improved by at least another order of magnitude at a dedicated muon EDM experiment at PSI
by employing the frozen-spin technique~\cite{Semertzidis:1999kv,Farley:2003wt,Adelmann:2010zz}. The corresponding sensitivities are illustrated in Fig.~\ref{fig:sensitivity}.
In particular, the dedicated muon EDM experiment would be able to probe most of the parameter space for which the current $(g-2)_\mu$ tension implies an $\Order(1)$ phase
of the Wilson coefficient.

\begin{figure}[t!]
	\centering
	\includegraphics[width=0.7\linewidth]{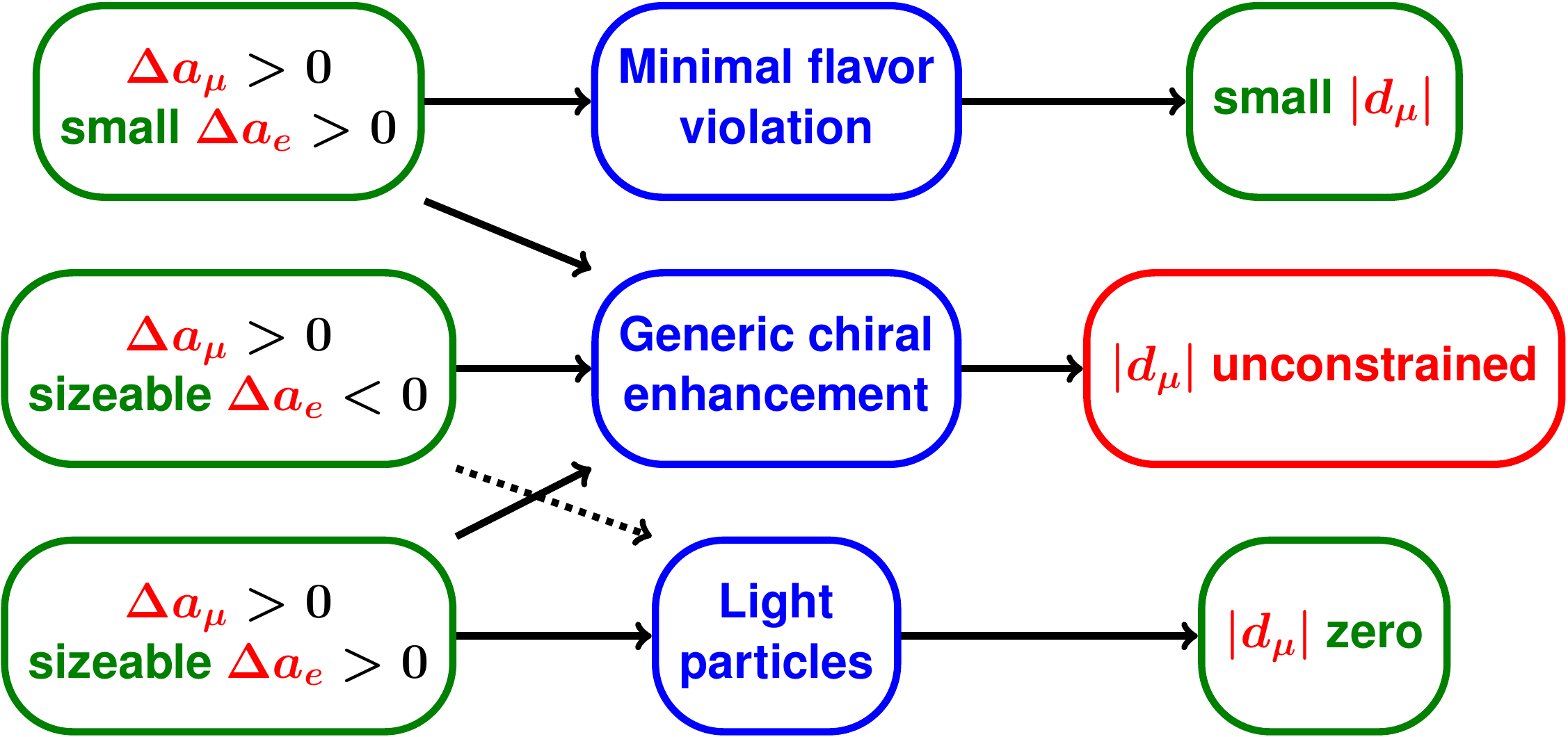}
	\caption{Possible scenarios for the $(g-2)_{\mu,e}$ deviations and implications for the muon EDM.}
	\label{fig:flowchart}
\end{figure}

The overall situation can thus be described as follows, see Fig.~\ref{fig:flowchart}:
\begin{enumerate}
 \item\label{scenario_1} A scenario in which both $\Delta a_\mu$ and $\Delta a_e$ are positive with a small $\Delta a_e$ is naturally accommodated in MFV. If MFV is realized, the resulting $|d_\mu|$ is strongly constrained by the limits on $|d_e|$, see Eq.~\eqref{dmu_MFV}, and out of reach experimentally.
 \item A scenario in which $\Delta a_\mu>0$ and $\Delta a_e<0$ and sizable, as indicated by the present experimental results in Eqs.~\eqref{Delta_amu} and~\eqref{Delta_ae},
 could be realized in models with chiral enhancement. In such cases $|d_\mu|$ is unconstrained, and the upcoming $g-2$ experiments, but especially a dedicated muon EDM experiment, would 
 probe a large portion of the parameter space corresponding to $\Order(1)$ phases.
 \item\label{scenario_3} A scenario in which both $\Delta a_\mu$ and $\Delta a_e$ are positive with a sizable $\Delta a_e$ could point towards an explanation in terms of light particles. In these models the muon EDM 
 vanishes because the Wilson coefficient is real.
\end{enumerate}
We stress that chiral enhancement could also occur in scenarios~\ref{scenario_1} and~\ref{scenario_3}, but in these cases would not offer an obvious advantage over alternative 
explanations within MFV or with light particles. In addition, as shown by the model constructed in Ref.~\cite{Davoudiasl:2018fbb}, the present situation can still be realized with light particles, 
exploiting an interplay between one- and two-loop diagrams, but this model is not yet UV complete, with one proposed completion again involving vector-like fermions.

We conclude that improved measurements of the muon EDM, especially in combination with the anticipated
progress for $a_e$, $a_\mu$, and $\alpha$, would provide valuable complementary insights and complete  
the search for BSM physics in lepton magnetic moments. If the current tensions were to persist, it would
help disentangle the flavor structure of the underlying BSM scenario.

\section*{Acknowledgments}

Financial support by the DOE (Grant No.\ DE-FG02-00ER41132) is gratefully acknowledged. AC is supported by a Professorship Grant (PP00P2\_176884) of the Swiss National Science Foundation. 

\end{document}